\documentclass{elsart}
\usepackage{graphicx}
\usepackage{natbib}
\usepackage{amsbsy}

\bibpunct{(}{)}{;}{a}{}{,} 

\newcommand{\be}{\begin{equation}}
\newcommand{\ee}{\end{equation}}
\newcommand{\bvt}{\begin{verbatim}}
\newcommand{\evt}{\end{verbatim}}
\newcommand{\bea}{\begin{eqnarray}}
\newcommand{\eea}{\end{eqnarray}}
\newcommand{\bc}{\begin{center}}
\newcommand{\ec}{\end{center}}

\begin{document}
\runauthor{Becciani, Antonuccio and Comparato}
\begin{frontmatter}
\title{FLY: MPI-2 High Resolution code for LSS Cosmological Simulations}
\author[OACT]{U. Becciani} 
\author[OACT]{V. Antonuccio-Delogu} and
\author[OACT]{M. Comparato}

\address[OACT]{Osservatorio Astrofisico di Catania,
              Via Santa Sofia 78, I-95123 Catania, ITALY\\
              e-mail: ube@oact.inaf.it
             }

\begin{abstract}
Cosmological simulations of structures and galaxies 
formations have played a fundamental role in the study of the origin,  formation and 
evolution of the Universe. These studies improved enormously with the use of 
supercomputers and parallel systems and, recently, grid based systems and Linux  clusters.
Now we present the new version of the tree N-body parallel code FLY that runs on a PC Linux Cluster using the 
one side communication paradigm MPI-2 and we show the performances obtained. 
FLY is included in the Computer Physics Communication Program Library.
This new version was developed  using the Linux Cluster of  CINECA, an IBM  Cluster  with 1024 
Intel Xeon Pentium IV 3.0 Ghz. The results show that it is possible to run a 64 Million particle simulation
in less than 15 minutes for each timestep, and the code scalability with the number of processors is achieved.
This lead us to propose FLY as a code to run very large N-Body simulations with more than $10^{9}$ particles  
with the higher resolution of a pure tree code. 
The FLY new version will be  available at  http://www.ct.astro.it/fly/ and CPC Program Library. 

\noindent 

\end{abstract}
\end{frontmatter}               

{\bf NEW VERSION PROGRAM SUMMARY}

\section{Introduction}
In cosmological applications, N-body codes are used to study the evolution of the structure formation
throughout the history of the Universe. In a simulation,  particles represent such large aggregates  that galaxies are typically just resolved. Simulations must follow a box large enough to
accurately represent the power spectrum of fluctuations on very large scales,
to compare them with real data. The number of particles then sets
the mass resolution of the simulation, which we would like to make as fine as possible. This
requires very large values of N, and state-of-the-art simulations follow up to 10 billion particles \cite{V.Springel2} \cite{Gao} and
codes that evolve both N-body particles and gas. 
FLY is a parallel tree code that runs with a very high resolution, N-Body simulation of the Large Scale Structure 
of the Universe, and could be integrated with a code that executes the hydrodynamic system evolution using a Paramesh structure \cite{Paramesh}  \cite{Comparato}.\\
Among the most adopted N-body codes for cosmological simulations there is the {\it Gadget} code \cite{V.Springel1}.
Gadget is a TreeSPH \cite{Hernquist} code for cosmological evolution, for simulations of  cosmological regions, 
considering both the collisionless matter (dark matter) and an ideal gas.  The gravitational interactions are computed 
using a tree algorithm (a hierarchical multipole expansion), while the gas dynamics uses a smoothed
particle hydrodynamics schema (SPH). Both gas and dark matter are  represented by particles. In the
new version of Gadget (GADGET-2) \cite{V.Springel} gravitational forces are computed with a hierarchical multipole expansion, which can optionally be applied in the form of a TreePM algorithm, where only short-range forces are computed with the tree-method while long-range forces are determined with Fourier techniques.\\
Another largely used code in cosmology is the {\it Enzo} code \cite{Bryan}. 
Enzo uses a totally different approach to collisionless systems. It allows the execution of hydrodynamic and N-Body
simulations using the adaptive mesh refinement technique \cite{Berger&Colella}. The dark matter particles are sampled 
in a grid structure to form a spatially discretized  field and  Poisson's equation of the dark matter evolution 
is solved by using the FFT method. The hydrodynamic part is solved by 
using a modified version of the piecewise parabolic method (PPM) \cite{Woodward&Colella}.
To conclude we cite the {\it PMFAST} code \cite{Merz} based on the MPI and OpenMP  paradigm. 
The forces are divided into short range components and  long range components. The short range components are
computed by using a fine mesh, the long range components are computed by using a coarse mesh (four times coarser than the fine mesh) and both use the FFT algorithm.
In this scenario FLY is a free parallel tree N-body code that allows researchers to run cosmological simulations
with higher resolution and a very high number of particles. Simulations with more than $10^8$ can be easily done
even using a small cluster. 
In the following sections we will describe the main  features and the obtained  performances in a Linux Cluster system.

\section{FLY code description}

FLY is a parallel tree N-body code for cosmological simulations of the Large Scale Structure 
of the Universe based on the Barnes-Hut algorithm \cite{BH}. 
The code is written in Fortran 90 and C, and it is based on the one-side 
communication paradigm. FLY creates the MPI Window object for one-side communication 
for all the shared arrays that
 can be accessed from
all  the processes,  avoiding any kind of synchronism. The code version 1.0 was
originally developed on CRAY T3E and  SGI ORIGIN systems using the logically SHared MEMory access routines 
({\it SHMEM}). The FLY version 2.1 was implemented for 
IBM SP by using the Low-Level Application Programming Interface routines ({\it LAPI}).\\
This new code (version 3.1) is a stable version that can run on a Linux platform, from a single PC
to a Linux Cluster. This is the evolution of preliminary codes and it reaches very high performance 
in all the systems where it has been tested. 
The main goal is to provide  researchers with a powerful code for cosmological simulation with higher resolution 
compared with other public domain codes.\\
The new version of FLY is implemented by using the MPI-2 standard. The first release was implemented 
in the IBM SP system, but a stable version was written by using the MPICH2 library  on a PC Linux
cluster, obtaining very good results (the FLY performance is reported in section 4).
MPICH2  provides a new MPI implementation designed to implement the MPI-2
additions: dynamic process management, one-side operations, parallel I/O and other extensions. They 
 provide a vehicle for MPI 
implementation research and to develop new and better parallel programming environments.
MPICH2 has a set of daemons (called mpd's) that verify the communication among  machines before running
 parallel processes. MPICH2 implements a portable {\bf mpiexec} command to start  parallel applications.

\subsection{Equations of motion}
A detailed discussion on the discretized equations of motion used in FLY can be found in the reference guide 
\cite{Antonuccio2003} paragraph 2.  Here we report a short summary of the equations.\\
The Friedmann-Robertson-Walker metric is characterized by an expansion factor $a(t)$, where $t$ is the conformal
time. Let $\boldsymbol{x}_{i}(t)$ be the comoving coordinate of the $i$-th particle and $m_{i}$ its mass, then the
equations of motion are given by:
\be 
\dot{\boldsymbol{x}_{i}} = \boldsymbol{v}_{i}  \label{eq1}
\ee 
\be
\dot{\boldsymbol{v}_{i}} + 2\frac{\dot{a}}{a}\boldsymbol{v}_{i} = - \frac{G}{a^{3}}
\sum_{j\ne i}\frac{m_{j}(\boldsymbol{x}_{i}-\boldsymbol{x}_{j})}
{\mid\boldsymbol{x}_{i}-\boldsymbol{x}_{j}\mid^{3}} +
\boldsymbol{F}_{Ewald}(\boldsymbol{x})  \label{eq2}
\ee

where $G$  is the gravitational constant, the term, $\boldsymbol{F}_{Ewald}(\boldsymbol{x})$  represents 
 {\em Ewald correction},
which takes into account the contribution to the force from the periodical boundary conditions, 
and the dot denotes the derivation compared to the conformal time $t$. We also define the {\em Hubble constant}: 
$H(t)=\dot{a}/a$.\\
It is more convenient to introduce a set of dimensionless variables:
\[
\boldsymbol{x}_{i}'=L_{0}\boldsymbol{x}_{i},\hspace*{0.5cm} t=t_{0}\tau,\hspace*{0.5cm}
m_{i}'=M_{0}m_{i}  
\]
In terms of these variables, the dimensionless equations of motion become:
\be
\frac{d\boldsymbol{x}_{i}'}{d\tau}=\boldsymbol{v}_{i}'  \label{eq3}
\ee
\be
\frac{d\boldsymbol{v}_{i}'}{d\tau}+2H(\tau)\boldsymbol{v}_{i}'=\frac{GM_{0}t_{0}^{2}}{L_{0}^{3}}
\frac{\boldsymbol{a}_{i}'}{a(\tau)^{3}}  \label{eq4}
\ee
 A detailed discussion on the measure units, the choice of the time variable, the
 adopted gravitational potential (in the Plummer form), the choice of the dynamic time stepping criterion and 
 the discretized equations used by the FLY code can be found in \cite{Antonuccio2003} paragraph 2. 

\subsection{FLY technical features}

The domain decomposition criterion is a fundamental point of all the cosmological codes. 
The data  distribution criterion must balance the load among the processors 
and  minimize the communication on the network.
The main data structures, particles and  tree cells, are statically divided 
among the processes to ensure a good initial balance of the load and 
to avoid any bottleneck while accessing remote data.\\
FLY assigns an equal number of particles to each process: running with $N_p$ processes each
process has an equal portion of the particles structure (i.e. pos(1:3,$N_{part}/{N_p}$).
Using this kind of assignment we developed a dynamic load balance procedure 
that makes use of the one-side communication characteristics, data grouping and data buffer.\\
Using the $FLY_{sort}$ utility (\cite{Becciani} paragraph 3),
a sorted  input file is obtained. It contains the fields 
of position and velocity, so that particles with a near 
tag number are also close in the physical space and located in the same process. 
Considering that the direct interaction among the nearest particles
has a relevant weight in the force computation, this distribution guarantees the minimization
of the communications and the maximization of the local computation.\\
The tree cells are numbered progressively from the root, down to the smallest cells. 
The optimal data distribution  is reached by using 
a fine grain data distribution.  More details are reported in a following paragraphs.\\
Another important feature consists in the grouping force calculation. The basic idea
 is to build a single interaction list
to be applied to all particles inside a {\it grouping cell} $C_{group}$ of the tree.
 This reduces the number of  tree access, and  builds a single  interaction list, that is the
 list of elements used to compute the force for each particle in the grouping cell \cite{Becciani2000}.
The last important feature consists in the data buffering. The data buffering uses all the free memory, 
not allocated to store arrays containing remote particles and the tree cells properties. The policy of
management of this structure is  based on the common management of a cache memory. 
Every time the processor  accesses a remote element, it first  looks for the local data buffer. 
If the element is not found, the process executes the GET calls to download the remote element and stores
it in the local data buffer. 

\subsection{Dynamic Load Balance} 

The Load Balancing and high performances are achieved by using  the above mentioned  grouping features and the 
one-side communication system  described in \cite{Becciani} and \cite{Becciani2000}, 
and hereafter shortly reported.\\
Each particle or grouping
 cell has an {\it executor processor} (hereafter PEx) that computes the force for the particles.
The PEx for a group is the processor where the main number of particles are stored. Equally, 
the PEx for a particle is the processor where the particle properties are stored. 
First of all, each PEx computes the  force  for all particles in the grouping cells. 
When a processor has no more groups to compute forces, it can start the force computing  phase for other $C_{group}$ 
cells, not yet
computed by the  default  PEx. In a similar way, FLY balances the load for the remaining particles; when a processor has no more local particles to compute forces, it can start this  phase for other particles, not yet
computed by the  default  PEx.\\
The one-side communication paradigm allows FLY to perform
this task without synchronism or waiting states among the PEs, and to obtain a high
load balance in this phase.\\
In this way each processor starts to work  on  local groups and particles, at the end it
can continue to work for remote groups and particles and only stop  when no more particles in the simulation
 need to be computed, avoiding the
load imbalance. 

\subsection{Data distribution}

The data distribution is described  in \cite{Becciani} and \cite{Becciani2000}, 
and hereafter shortly reported.\\
The tree cells are numbered progressively from the root,  down to the smallest cells. 
A good data distribution scheme is reached by using 
a fine grain data distribution as reported in the above mentioned articles. The first 
tree levels contain cells that  are always checked to form the list of cells and particles needed to compute forces
for a fixed body.\\
This data distribution presents all the processors from investigating the same cells in the same remote memory and avoids the typical
problems of {\it access to a critical resource}: a tree fine grain data distribution allows, on
average, all the PEs memories to be requested with the same frequency; thus each particle 
will have the same average access time to the tree cells,  avoiding the bottleneck problem.\\
Particle properties are  organized with the following schema. Each 
processor has the same number of particles, $Nbodies/N\_processors$, near in  space, using the sort
utilities as described in  \cite{Becciani}. This kind of distribution in contiguous blocks,
 is a good data distribution in terms of measured code performance.The list of particles we need to compute forces often includes near particles that are locally stored, and the communication is minimized.
 
\section{The MPI-2 code version}

FLY is based on the one-side communication paradigm. The new version adopts the MPICH2 library. The
main data structure, that has access from remote processes, is declared in a module procedure of FLY (fly\_h.F90
routine).  Then FLY creates the MPI Window object for one-side communication for all the shared arrays,
with a call like the following:
\begin{verbatim}
CALL MPI_WIN_CREATE(pos, size, real8, MPI_INFO_NULL,
                    MPI_COMM_WORLD,win_pos,ierr)
\end{verbatim}
the following main window objects are created: 
\begin{itemize}
\item win\_pos, win\_vel, win\_acc:  particles positions velocities and accelerations
\item win\_pos\_cell, win\_mass\_cell, win\_quad, win\_subp, win\_grouping: cells positions, masses, quadrupole momenta, tree structure and grouping cells.  
\end{itemize}
Other windows are created for dynamic load balance and global counters.\\
The main phases where  communication occurs are the following: the tree construction phase 
and the force computation.\\
During the tree construction phase, all the precesses cooperate to build the single tree 
structure of the simulation. In the {\it tree\_gen.F90} routine, each process mainly
computes the cells that are locally resident. The tree is  built level by level, one cycle
for each level. Every level subdivides the cells  into 8 sub-cells and prepares them for the new level. 
The processes compute the number of particles in each sub-cell: sub-cells with more than one particle form the
cells of the new level. During this phase FLY must access to remote cells. 
Using the data buffering the window locking calls are always of shared type, and the  {\it put} operation, 
like the following, is often required 
\begin{verbatim}
CALL MPI_WIN_LOCK(MPI_LOCK_SHARED, ind_pe_rmt, 0,
                  win_subp, ierror) 
CALL MPI_PUT(subp_ch(K,J), 1, MPI_INTEGER4, ind_pe_rmt,
             startIndex, 1, MPI_INTEGER4, win_subp, ierror)
CALL MPI_WIN_UNLOCK(ind_pe_rmt, win_subp, ierror)
\end{verbatim}

Sometimes, depending on the data buffer dimension, the {\it accumulate} operation, like the following, can occur

\begin{verbatim}
CALL MPI_WIN_LOCK(MPI_LOCK_SHARED, ind_pe_rmt, 0,
                  win_subp, ierror) 
CALL MPI_ACCUMULATE(subp_ch(K, ind_ch), 1, MPI_INTEGER4,
                    ind_pe_rmt,startIndex, 1, MPI_INTEGER4, 
	            MPI_SUM, win_subp, ierror)
CALL MPI_WIN_UNLOCK(ind_pe_rmt, win_subp, ierror)
\end{verbatim}

During the force computation phase of the Barnes-Hut algorithm, FLY mainly uses the get procedure: 
generally speaking, it reads remote bodies and cell properties in a tree walk procedure, and computes the
force for locally residing bodies.
\begin{verbatim}
CALL MPI_WIN_LOCK(MPI_LOCK_SHARED, ind_pe_rmt, 0,
                  win_pos, ierror)
CALL MPI_GET(pos_cell(1), ndim, MPI_REAL8, ind_pe_rmt,
             startIndex, ndim, MPI_REAL8, win_pos, ierror)
CALL MPI_GET(pmass(nterms), 1, MPI_REAL8, ind_pe_rmt,
             startIndex, 1, MPI_REAL8, win_mass_cell, ierror)
CALL MPI_GET(pquad(1, nterms), 5, MPI_REAL8, ind_pe_rmt,
             startIndex, 5, MPI_REAL8, win_quad, ierror)
CALL MPI_WIN_UNLOCK(ind_pe_rmt, win_pos, ierror)
\end{verbatim}

FLY verifies which remote elements (cells or bodies) must  be considered 
to compute the force on a given particle, and gets the remote data. Local data are obviously accessed
without  MPI communication.
All the accesses in this phase are carried out with a LOCK\_SHARED access. For the dynamic load balance and the grouping 
computation, there are few LOCK\_EXCLUSIVE calls in critical sections, mainly to update global counters in
the {\it acc\_comp.F90} routine.\\
At the end each process updates the particle position and a new timestep is started.
The code has totally more than 570 MPI calls. More than 110 calls are for the  MPI\_WIN\_LOCK and UNLOCK windows in 
shared mode to perform get and put operations and few FENCE calls. All the GET operations occur during a phase when 
data in the windows are not updated and the PUT operation occurs mainly in physical locations that are accessed by any 
process. Only few exclusive locks are required. About 60 MPI\_GET and PUT operations are required and few global 
counters are implemented using  MPI\_ACCUMULATE calls.

\section{FLY performance}

The FLY MPI2 version was developed on the IBM Linux Cluster at Cineca, with the following feature
Architecture: IBM Linux Cluster 1350, 512 nodes with 2 Processors for each node and 2GB Ram for each processor.
Processor type: Intel Xeon Pentium IV 3.0 Ghz and 512 KB cache (128 nodes have  Nocona 
processors). Internal Network: Myricom LAN Card "C" Version and "D" Version. Operating System: Linux SuSE 
SLES 8.\\
The code was compiled using the mpif90 compiler version 8.1 and with basic  optimization options in order to have performances that could be useful compared with other generic clusters. We use the following compilation options:
\begin{verbatim}
mpif90 -O3 -tpp7  -static -xN
\end{verbatim}
where O3 enables aggressive optimizations, tpp7 optimizes the code for Pentium IV processor, static avoids linking with shared libraries and xN generates a specialized code to run exclusively on  Intel Pentium IV processors and compatible Intel processors. The mpich2 1.0 was installed and used for these tests, and  native communication protocol was also used.\\
We run FLY using a cosmological  CDM$+\Lambda$ model ($\Omega=0.3$, $\lambda=0.7$, $h=0.6$)  with a different number of
particles in order to test the scalability of the code in the Intel cluster and the scalability of the system. The following paragraph reports the results  obtained.

\subsection{Scalability}
\begin{figure}
\centering
\includegraphics[width=9cm]{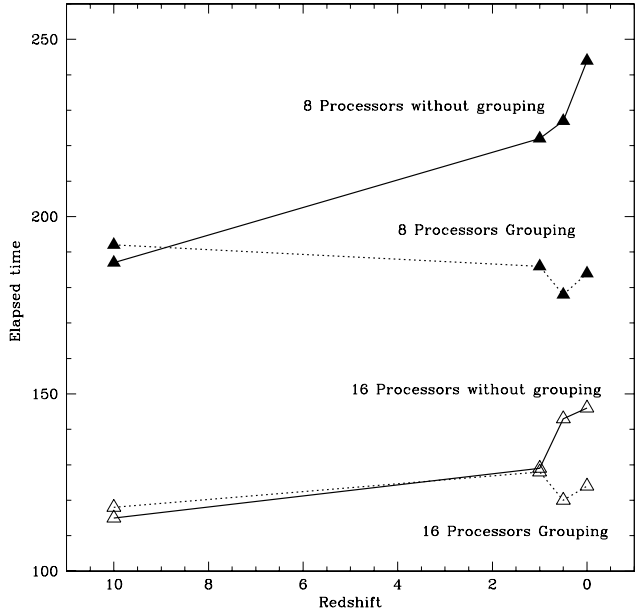}
\caption{Elapsed time versus redshift of a 2 Million particle simulation}
\label{fg:red}
\end{figure}

In this section we report scalability data  using two  testcases with  2 Million and 16 Million particles, 
in a uniform initial condition (z=80). Even if the timestep could be twice slower when clusters form, the grouping features of FLY avoid this behaviour and a single timestep duration will not increase or will increase  by no more than $20$\%.
Initial data were generated using a tool based on COSMICS \cite{Cosmics},
that is a package  for computing transfer functions and microwave background anisotropy; it  also generates gaussian random initial conditions for nonlinear structure formation simulations. We modified the original package to obtain an output format directly used by FLY. We generated a normalized  matter power spectrum distribution and constrained random density fields on a lattice, using the Hoffman-Riback algorithm \cite{Hoff}.
\begin{figure}
\centering
\includegraphics[width=9cm]{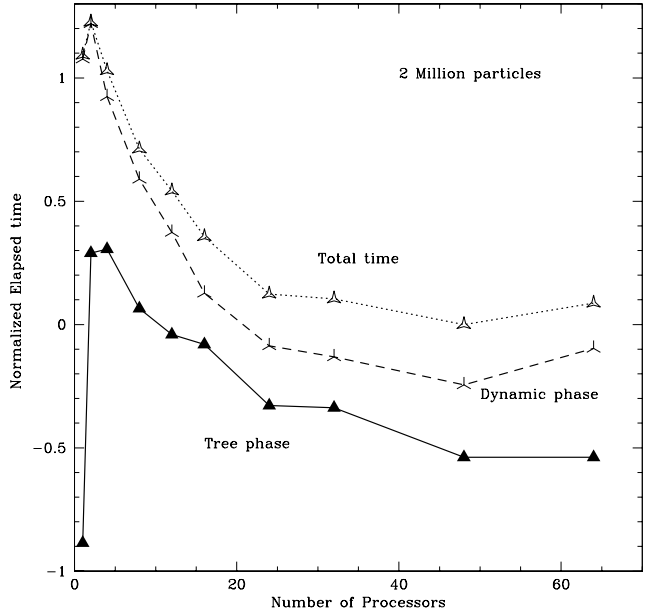}
\caption{Timing of FLY phases. Data represent the logarithm value of each FLY phase, normalized to the Total time obtained using 48 Processors}
\label{fg:sec2m}
\end{figure}
\begin{table}
\centering
\begin{tabular} {|c|r|r|r|} \hline
\centering

Processors	& Tree-phase & Dynamic-phase 	& T-step 	 \\  \cline{1-4}
1 &	5.50 &	494.52 &  512.45 \\ 
2 &	80.80 &	686.79 &  774.53 \\ 
4 &	83.90 &	348.36 &  444.10  \\
8 &	48.19 &	160.91 & 213.49	  \\ 	
12 &	37.92 &	97.93 &	144.17 \\ 
16 &	34.37 &	55.36 &	 94.12 \\ 
24 &	19.56 &	 33.85 & 55.26 \\ 
32 &	19.13 &	30.81 &	52.74 \\ 
48 &	12.38 &	23.65 &	41.45  \\ 
64 &	12.00 & 33.06 &	50.72  \\ 
\hline

\end{tabular}
\\
\caption{Elapsed time running  2 Million particles}
\end{table}
Fig. \ref{fg:sec2m} shows a graphical representation of the code scalability with 2 Million particles, increasing the number of processors. Data are normalized to the total timestep duration using 48 processors (41.5 seconds) and the logarithm value is plotted.   Table 1 displays the elapsed time in seconds of  each main phase of the FLY run. 
A single timestep of FLY is mainly for the tree construction phase and for the dynamic evolution of particles.
We report the tree and the dynamic phases together with the total time of a single timestep. This result shows that, in this case, it is not  very useful to make this simulation with more than 48 processors. \\
There is an important behaviour of the system running with one processor and with two or more processors: the tree phase increases by one order of magnitude. The result measured in the serial run, regards a tree located in the same local memory of the processor that runs the application.
The tree construction phase-time in a parallel run, mainly depends on  the atomic operation performed to build the tree, that is shared among the processors. All the processors cooperate to build the tree,  shared in the global memory, and they must manage shared counters to perform this task \cite{Antonuccio2003}. It is not possible to run FLY with two or more processors, without building the tree in a parallel way: the parallel run can take place only if the tree is shared among the processors memory. 
The code globally has a good scalability, and in particular the dynamic phase, whereas the percentage of the tree part ranges only from $10$\%   to $35$\% of the timestep duration. 
\begin{figure}
\centering
\includegraphics[width=9cm]{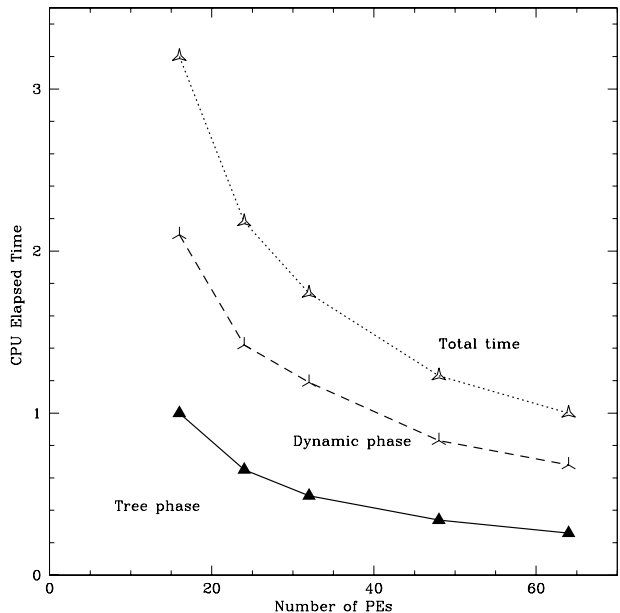}
\caption{Elapsed time of a single timestep with 64 Million particles}
\label{fg:sec64m}
\end{figure}
\begin{table}
\centering
\begin{tabular} {|c|r|r|r|} \hline

Processors	& Tree-phase & Dynamic-phase 	& T-step 	 \\  \cline{1-4}
16 &	828.56 & 1728.63 & 2630.98 \\ 
24 &	538.17 & 1166.43 & 1790.89 \\ 
32 &	399.02 & 978.20 & 1427.42 \\ 
48 &	276.61 & 686.49 & 1015.41 \\ 
64 &	217.01 & 556.93 &   822.64\\ 
\hline

\end{tabular}
\\

\caption{Elapsed time running  64 Million particles}
\end{table}
Fig. \ref{fg:sec64m} shows a graphical representation of  the code scalability with 64 Million particles, increasing the number of processors. Data are normalized to the total timestep duration using 64 processors (822.64 seconds).   Table 2 displays the elapsed time, in seconds, of  each main phase of this case. 
The results of this run show that the scalability for large LSS simulations in this system architecture is very good. They allow us to have  the same performance than using typical MPP systems. 
Fig. \ref{fg:sec64m} shows the results starting from $16$ processors because it is not possible to execute this simulation with less than 32 GB Ram.
\begin{figure}
\centering
\includegraphics[width=9cm]{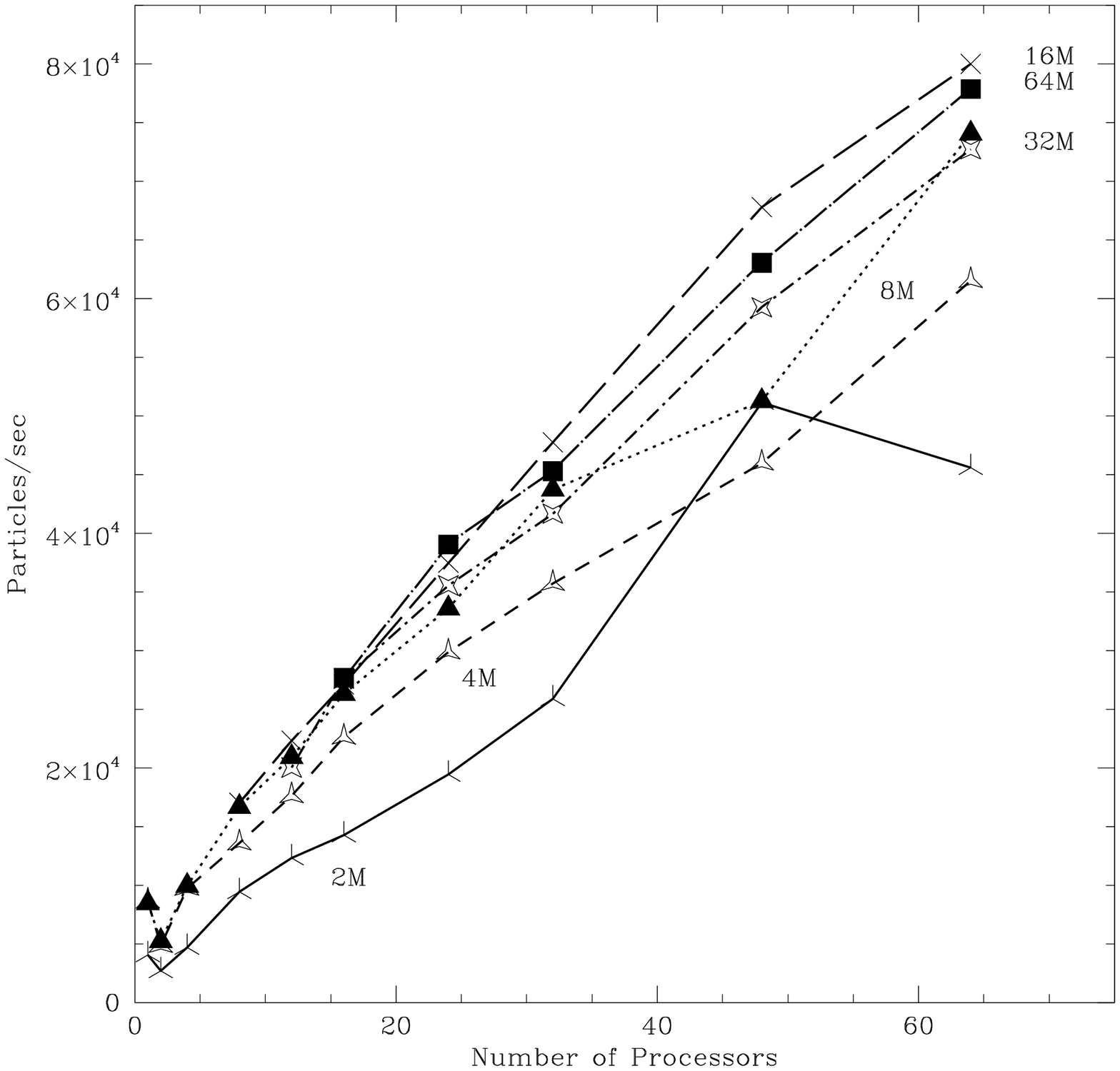}
\caption{Number of particles computed  each second increasing the size of a simulation}
\label{fg:pps}
\end{figure}
Fig. \ref{fg:pps} displays us the global results in terms of code scalability for this kind of architecture. The results of the simulation, running with 2 Million particles, show that with more than 48 processors, the system reaches a performance saturation, whereas a good accordance and a good scalability running with more than 48 Million particles is reached.
The user that wants  to run FLY in similar systems must consider the above results as a reference case. 
All data mentioned in the previous figures are considered at the beginning of a simulation (redshift $80$, in our case) when the particles are in a uniform distribution. 
The FLY code grouping working mode allow the user to set a grouping factor so that the error is much lower than the tree schema and can be  negligible. In this case the elapsed time for each time-step does not increase. 
During the evolution at the beginning there are no cluster formations and the elapsed time for each time-step is roughly constant. When the simulation starts to form clusters, FLY groups also start  to form, as mentioned in section 2.2, and this reduces the time-step duration. More details can be found in \cite{Becciani2000}. Fig. \ref{fg:red} reports data of a simulation of 2 Million particles without FLY groups and with a grouping factor of "level 7" with no more than 16 particles in each grouping cell. In this case at z=10 only   164367 particles were grouped and the grouping effect was negligible, but at the redshift z=0 more than 1.2 Million particles were grouped. A similar behaviour is obtained increasing the number of processors and/or the particles of a simulation.

\subsection{Isogranularity}
An important test is also given to allow us to make some considerations on the performance of this kind of architecture for the FLY code. 
We measured that a run with 16 Million particles using 16 Processors, each having 2 GB Ram, produces about $1.6 \cdot 10^{10}$ remote operations, mainly data GET and data PUT. FLY uses the data buffer  as described in \cite{Becciani},  storing remote data in a local buffer that is managed  as in the cache memory.\\
\begin{figure}
\centering
\includegraphics[width=9cm]{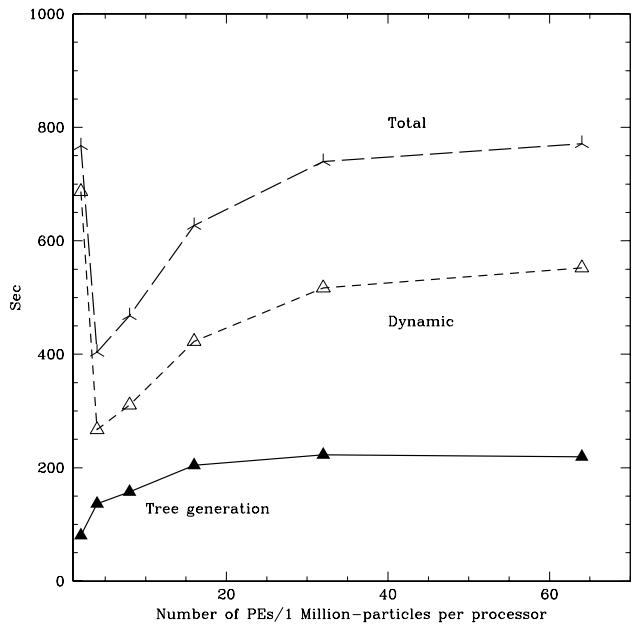}
\caption{Elapsed time increasing size of a simulation}
\label{fg:iso}
\end{figure}
Fig. \ref{fg:iso} reports the behaviour of the system increasing the size of a simulation as the number of processors and of the global RAM grows. The base case was to run 1 million particles for each processor with 2 GB Ram. We test the parallel case only and the  curve in the figure starts from 2 Million particles running in two processors.
The total timestep and the dynamical part of the code show that using two processors in the same node we do not obtain the best performances, due to the intra-node network contention for the access to critical resources. With more than two processors we measured an increased elapsed time with the number of processors; the total elapsed time with 64 processors is twice  as 4 processors, this behaviour depending on the network contention of the system.
in

\section{Testcase Cosmological simulation}
This section gives a short description of an example of a simple FLY run with the aim to provide a testcase the user can execute. FLY is included in the CPC program library. The testcase we describe can be downloaded from the FLY page, at http://www.ct.astro.it/fly/. \\
We run a 2 Million particle simulation. The initial conditions were created using Cosmics \cite{Cosmics} in a $50 h^{-1} {\rm Mpc}$ cubic region, within a  CDM$+\Lambda$ cosmogeny corresponding to $\Omega_0=0.3, \Omega_{lamda}=0.7, zstart=80$, and $h=0.6$. FLY implements a set of cosmological equations of motion, solving the standard particle
equations of motion for a Friedmann cosmology, with the {\em Ewald correction},
which takes into account the contribution to the force from the infinite replicas of the
simulation box over the spatial directions. All the parameters reported in the testcase are discussed in \cite{Antonuccio2003}.\\
The user must download FLY source code and compile it using the parameter $nknots=49$ in the {\it fly\_h.F90} module file and files provided in the link testcase.\\
We executed the simulation in the Cineca Linux cluster  mentioned above, with 32 processors. The simulation evolved without grouping factor for 66 timesteps in a total time of about $100$ minutes and using the native communication protocol.

\section{Conclusions}
This new free release of FLY (version 3.1) will give a contribution to the astrophysical community for two main new features, that can give new  opportunities and new results in the cosmological field.\\
It is now possible to execute  LSS cosmological simulations using FLY, a tree N-body code, with high resolution, using a Linux Cluster with the MPICH2 library. Moreover, FLY has a new interface to the code that can communicate data using a Paramesh like structure, giving researchers  new possibilities. It is possible to run two separate codes, both with  high resolution, using the same computational domain.\\
This code interoperability will be also exploited in the grid environment, specifying a devoted node to run FLY and another node to execute a fluid dynamic code. This new possibility will be considered in a new project to be developed at the INAF of Catania in the next few years.
 
\section{Acknowledgements}
All the tests carried out on the IBM Cluster system at the CINECA, were executed using the financial support
 of the INAF, the Italian National Institute for Astronomy. 
 We gratefully acknowledge the FLASH group  and in particular  Dr. T. Plewa  and Dr. D. Sheeler of the  ASCI/Alliances Center for Thermonuclear Astrophysical Flashes,  University of Chicago, for the development of the FLY interface towards the Paramesh based code \cite{Antonuccio06}, and Dr. L. Santagati of INAF - Catania Astrophysical Observatory for the english revision of the text. 




\begin{thebibliography}{999}
\bibitem{Antonuccio2003} Antonuccio, V., Becciani, U. and Ferro, D. {\em Comp. Phys. Comm.} {\bf 155} (2003) 159Antonuccio06
\bibitem{Antonuccio06} Antonuccio, V., Becciani, U., Plewa, T., Sheeler, D.  {\em Cosmological simulation with FLY-FLASH} (in preparation)
\bibitem{BH} Barnes, J. and Hut, P. {\em Nature} {\bf 324} (1986) 446
\bibitem{Becciani2000} Becciani, U. and Antonuccio-Delogu, V. {\em J. Comput. Phys.} {\bf 163} (2000) 118
\bibitem{Becciani} Becciani, U. and Antonuccio, V. {\em Comp. Phys. Com.} {\bf 136} (2001) 54
\bibitem{Berger&Colella} Berger, M. J. and Colella, P. {\em J. Comput. Phys.} {\bf 82} (1989)  64
\bibitem{Cosmics} Bertschinger, E. {\em Astro-Ph/9506070} (1995)
\bibitem{Bryan} Bryan, G. L. and Norman, M. L. {\em ASP Conf. Ser.} (1997) {\bf 123} 363
\bibitem{Comparato} Comparato, M., Antonuccio, V. and Becciani, U. {\em Mem. Sait}  (in press)
\bibitem{Gao}  {\em MNRAS} {\bf 363} (2005) L66
\bibitem{Hernquist} Hernquist, L.and Katz, N. {\em ApJS} (1989) {\bf 70} 419
\bibitem{Hoff} Hoffman, Y., Ribak, E.  {\em ApJ} {\bf 396} (1992) 448T
\bibitem{Paramesh} MacNiece, P., Olson, K., Mobarry, M. C.,  De Fainchtein, R. and Packer, C. {\em Comp. Phys. Comm.} {\bf 126} (2000) 330
\bibitem{Merz} Merz, U.  Pen, U. and Trac, H. {\em New Astron.} {\bf 10} (2005) 393
\bibitem{V.Springel} Springel,V., Yoshida, N. and White, S.  {\em New Astron.} {\bf 6} (2001) 79 
\bibitem{V.Springel1} Springel,V.  {\em MNRAS} {\bf 364} (2005) 1105
\bibitem{V.Springel2} Springel,V., Simon D.M., W. et al.  {\em Nature} {\bf 435, Issue 7042} (2005)  629
\bibitem{Woodward&Colella} Woodward, P. R. and Colella, P.{\em J. Comput. Phys.} {\bf 54} (1984)  115.
\end{thebibliography}
\end{document}